\documentclass[preprint,pre,nofootinbib,showpacs,titlepage]{revtex4}%
\usepackage{amsfonts}
\usepackage{amsmath}
\usepackage{amssymb}
\usepackage{graphicx}
\usepackage{acronym}%
\begin{document}
\title{Clustering and the Synchronization of Oscillator Networks.}
\author{Patrick N. McGraw and Michael Menzinger}
\affiliation{Department of Chemistry, University of Toronto, Toronto, Ontario, Canada M5S 3H6}

\begin{abstract}
By manipulating the clustering coefficient of a network without changing its
degree distribution, we examine the effect of clustering on the
synchronization of phase oscillators on networks with Poisson and scale-free
degree distributions. \ For both types of network, increased clustering
hinders global synchronization as the network splits into dynamical clusters
that oscillate at different frequencies. \ Surprisingly, in scale-free
networks, clustering promotes the synchronization of the most connected nodes
(hubs) even though it inhibits global synchronization. As a result, they show
an additional, advanced transition instead of a single synchronization
threshold.\ \ This cluster-enhanced synchronization of hubs may be relevant to
the brain which is scale-free and highly clustered.

\end{abstract}
\maketitle

A key problem in the study of networks is the relation between network
structure and function\cite{Strogatz}\cite{Netreviews}. Among the attributes
most frequently used to characterize network structure are degree distribution
and clustering coefficient. \ The latter, defined as the number of triangles
divided by the number of connected triples\cite{WS}\cite{Watts}, quantifies
the tendency of neighbors of a node to be also neighbors of each other. We
refer to this property as \emph{structural clustering} to distinguish it from
\emph{frequency }or \emph{dynamical clustering }\cite{Manrubia}%
\cite{Dyncluster}\cite{JalanAmritkar}.

A recent algorithm\cite{Kim} allows one to manipulate the clustering
coefficient by re-wiring the network, without changing its degree
distribution. Kim used this technique to study the effect of clustering on the
performance of variously structured Hopfield networks\cite{Kim}, and we use it
here to study its effect on synchronization.

\ In search of factors that control the synchronization of networks of
oscillators, researchers have studied the effects of different structural and
statistical attributes\cite{Manrubia}-\cite{Hwang}. These studies have
included systems of coupled maps\cite{Manrubia}\cite{JalanAmritkar}%
\cite{Gade}\cite{Jost}, continuous-time chaotic oscillators\cite{Barahona}%
\cite{WangChenSF}\cite{WangChenSW}\cite{Nishikawa}, spiking
neurons\cite{LagoFernandez}, and phase oscillators\cite{Dyncluster}%
\cite{Hong1}\cite{Moreno}\cite{Katriel}. \ Unsurprisingly, shortcuts in
small-world networks tend to improve synchronization compared to regular
lattices\cite{Watts}\cite{Gade}\cite{Barahona}\cite{LagoFernandez}%
\cite{Hong1}. However, other factors such as degree
heterogeneity\cite{Nishikawa}, maximum betweenness centrality\cite{Hong},
asymmetry and weighting of couplings\cite{Motter}\cite{Hwang} also play a role.

In the present work we used Kim's procedure\cite{Kim} to study the effects of
structural clustering on the synchronization of networks of
Kuramoto-like\cite{Kuramoto} phase oscillators with Poisson and scale-free
degree distributions. \ For both network types we found that increased
clustering impedes global synchronization and magnifies the fluctuations of
the global order parameter. Clustering also qualitatively changes the onset of
synchronization. At low clustering\cite{Hong1}\cite{Moreno}, the transition is
similar to that in the mean-field or globally coupled Kuramoto model, where a
single subset of oscillators becomes entrained at a central frequency. With
increasing coupling, this synchronized subset entrains larger portions of the
remaining oscillators. At higher clustering, on the other hand,
synchronization begins with more than one subset, each synchronized at a
different frequency, and continues at increased coupling through the
recruitment of remaining oscillators by the different subsets. \ We refer to
these subsets as \emph{frequency clusters}. In contrast to Poisson networks,
characterized by a single synchronization transition, we find that
\emph{strongly clustered scale-free networks show a second synchronization
transition} which is advanced relative to the principal transition, i.e. it
occurs at a lower value of the coupling strength. Thus, structural clustering
promotes synchronization at low coupling, while\ inhibiting it at higher
coupling. The advanced synchronization begins with the hubs, or highest-degree
nodes. \ These results appear to be relevant to natural networks, such as the
brain\ \ \cite{equiluz}\ \ and ganglia \cite{Watts}\cite{Netreviews}%
\cite{cherniak} that have higher clustering coefficients than those predicted
by simple growth models. In the brain the distribution of \ functional
connections and the probability of finding a link vs. distance are both scale
free \cite{equiluz}.

In contrast to approaches\cite{Barahona}\cite{WangChenSF}\cite{Nishikawa}%
\cite{Hong} based on the master stability function\cite{PecoraMSF}, we
consider non-identical oscillators and their full dynamics, including states
where only some of the oscillators are synchronized.

We consider networks of oscillators obeying the coupled differential equations%
\begin{equation}
\frac{d\phi_{i}}{dt}=\omega_{i}+\frac{\lambda}{\left\langle k\right\rangle }%
%TCIMACRO{\dsum \limits_{j}}%
%BeginExpansion
{\displaystyle\sum\limits_{j}}
%EndExpansion
a_{ij}\sin(\phi_{i}-\phi_{j}),\label{Kuramotomodel}%
\end{equation}
where $0\leq\phi_{i}<2\pi$ are $N$ phase variables, \ $\omega_{i}$ are the
randomly and uniformly distributed intrinsic frequencies, $\lambda$ is the
coupling strength, and $a_{ij}$ is the adjacency matrix (1 if $i$ and $j$ are
connected, 0 otherwise). \ The coupling strength is scaled by the average
degree $\left\langle k\right\rangle $ of the nodes, averaged over the whole
network. \ \ In the globally coupled Kuramoto model\cite{Kuramoto} where
$a_{ij}=1,\forall i\neq j$, this reduces to scaling by the number of
oscillators $N.$ \ The above form makes the coupling strengths symmetric and
weights all links equally\cite{Moreno}, but in the case of a nonhomogeneous
degree distribution, some oscillators may receive a stronger synchronizing
signal because they have more neighbors. \ 

Results are for networks with $N=1000$ nodes and average degree $\left\langle
k\right\rangle \approx20$. \ Qualitatively similar results were obtained with
$N=5000$ and $\langle k\rangle=6$.\ We consider a random (Poisson) network and
a scale-free network generated by the Barabasi-Albert preferential attachment
algorithm\cite{BA}, and families of networks derived from each of these by
changing the clustering coefficient $\gamma$. \ To vary $\gamma$, we use a
stochastic rewiring algorithm\cite{Kim} that rearranges connections with a
bias toward increased clustering. \ The procedure is as follows: \ 1) \ Pick
randomly two existing links. \ 2) Compute whether interchanging these links
increases or decreases the total number of triangles. \ Perform the
interchange only if it increases that number. 3) \ Repeat these steps until
the desired clustering is achieved. \ \ (One can reduce the clustering
coefficient by reversing the acceptance criterion). \ Since this algorithm
only rewires connections and does not change the degree of any node, the
degree distribution as well as the degree sequence is fixed.

The random $\omega_{i}$ values were uniform distributed over the interval
$0.9\leq\omega_{i}\leq1.1$ and the initial phases were also random. \ \ We ran
the dynamics at a series of increasing values of $\lambda$, \ integrating the
equations using a simple Euler method with step size $0.02$. \ After ~100 time
units to allow for relaxation to a steady state, \ we measured the
time-averaged (over 500 time units) synchronization order
parameter\cite{Kuramoto}%
\begin{equation}
m\equiv\left\{  \left\vert \frac{1}{N}\sum_{j=1}^{N}e^{i\phi_{j}}\right\vert
\right\}  .\label{orderparameterdef}%
\end{equation}
\ $m$ is of order $1/\sqrt{N}$ if the oscillators are uncorrelated and
approaches $1$ when all are in phase. \ In addition, we measured the frequency
$\Omega_{i}\equiv\left\{  \dot{\phi}_{i}\right\}  $of each oscillator over the
same 500-unit time interval after relaxation. The brackets $\left\{
{}\right\}  $ signify time averaging. \ \ This measurement reveals the
collective behavior in detail.

\begin{figure}
[ptb]
\begin{center}
\includegraphics[
height=4.1675in,
width=3.9427in
]%
{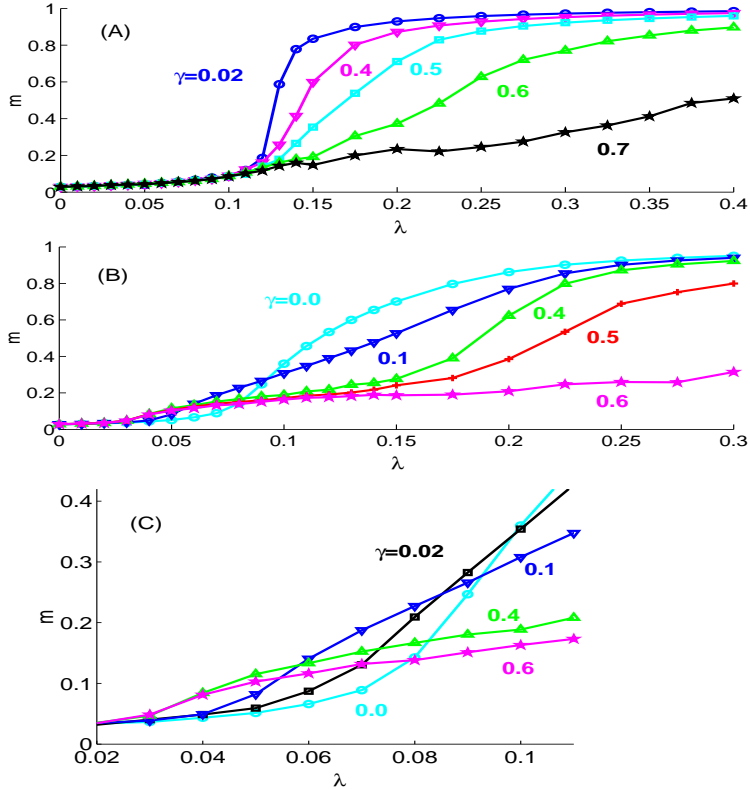}%
\caption{(color online) Order parameter $m$ vs. coupling strength $\lambda$, for different
values of the clustering coefficient $\gamma$. \ (A) \ Poisson degree
distribution. \ \ (B,C) Scale-free degree distribution. \ \ (C) Close-up of
the transition region showing that increased clustering leads to an advanced
(lower-$\lambda$) transition. \ \ }%
\label{syncplotboth}%
\end{center}
\end{figure}

To study the effect of clustering on the transition to synchronization, we
first examine the order parameter $m$ as a function of the coupling strength
$\lambda$. \ Figure \ref{syncplotboth}\ shows plots of $m$ vs. $\lambda$ for
networks with Poisson and scale-free degree distributions at several values of
the clustering coefficient $\gamma$. \ In each plot the data were averaged
over time, over several realizations of the intrinsic frequencies and over
several network rearrangements. \ The degree sequence, however, was the same
in all cases. \ Poisson networks (fig.1A) behave more simply than scale-free
ones. \ The natural (low-clustering) random network shows a rapid transition
to order above $\lambda=0.12$. \ \ Increasing clustering inhibits global
synchronization and makes this transition less steep. \ The lowered mean
values of $m$ in the highly clustered networks are associated with extremely
large temporal fluctuations around the mean. \ \ \ \
%TCIMACRO{\FRAME{ftbpFU}{3.9427in}{4.1675in}{0pt}{\Qcb{Order parameter $m$ vs.
%coupling strength $\lambda$, for different values of the clustering
%coefficient $\gamma$. \ (A) \ Poisson degree distribution. \ \ (B,C)
%Scale-free degree distribution. \ \ (C) Close-up of the transition region
%showing that increased clustering leads to an advanced (lower-$\lambda$)
%transition. \ \ }}{\Qlb{syncplotboth}}{syncplotboth.eps}%
%{\special{ language "Scientific Word";  type "GRAPHIC";  display "USEDEF";
%valid_file "F";  width 3.9427in;  height 4.1675in;  depth 0pt;
%original-width 6.7914in;  original-height 5.5486in;  cropleft "0";
%filename 'syncplotboth.eps';file-properties "XNPEU";}}}%
%BeginExpansion
%EndExpansion%
%TCIMACRO{\FRAME{ftbpFU}{6.3235in}{6.0226in}{0pt}{\Qcb{Poisson networks:
%Scatter plots of average oscillation frequency $\Omega$ (vertical axis) vs.
%intrinsic frequency $\omega$ (horizontal). \ At zero coupling, \ the points
%all lie on the line $\Omega=\omega$. \ With increasing coupling, \ more
%oscillators line up at synchronized frequencies. \ For high clustering,
%\ several bands form at different frequencies, rather than one frequency as in
%the low-$\gamma$ case.}}{\Qlb{poissonscatter}}{poissonscatter.eps}%
%{\special{ language "Scientific Word";  type "GRAPHIC";  display "ICON";
%valid_file "F";  width 6.3235in;  height 6.0226in;  depth 0pt;
%original-width 6.5838in;  original-height 5.2157in;  cropleft "0";
%filename 'poissonscatter.eps';file-properties "XNPEU";}}}%
%BeginExpansion
\begin{figure}
[ptbptb]
\begin{center}
\includegraphics[
height=6.0226in,
width=6.3235in
]%
{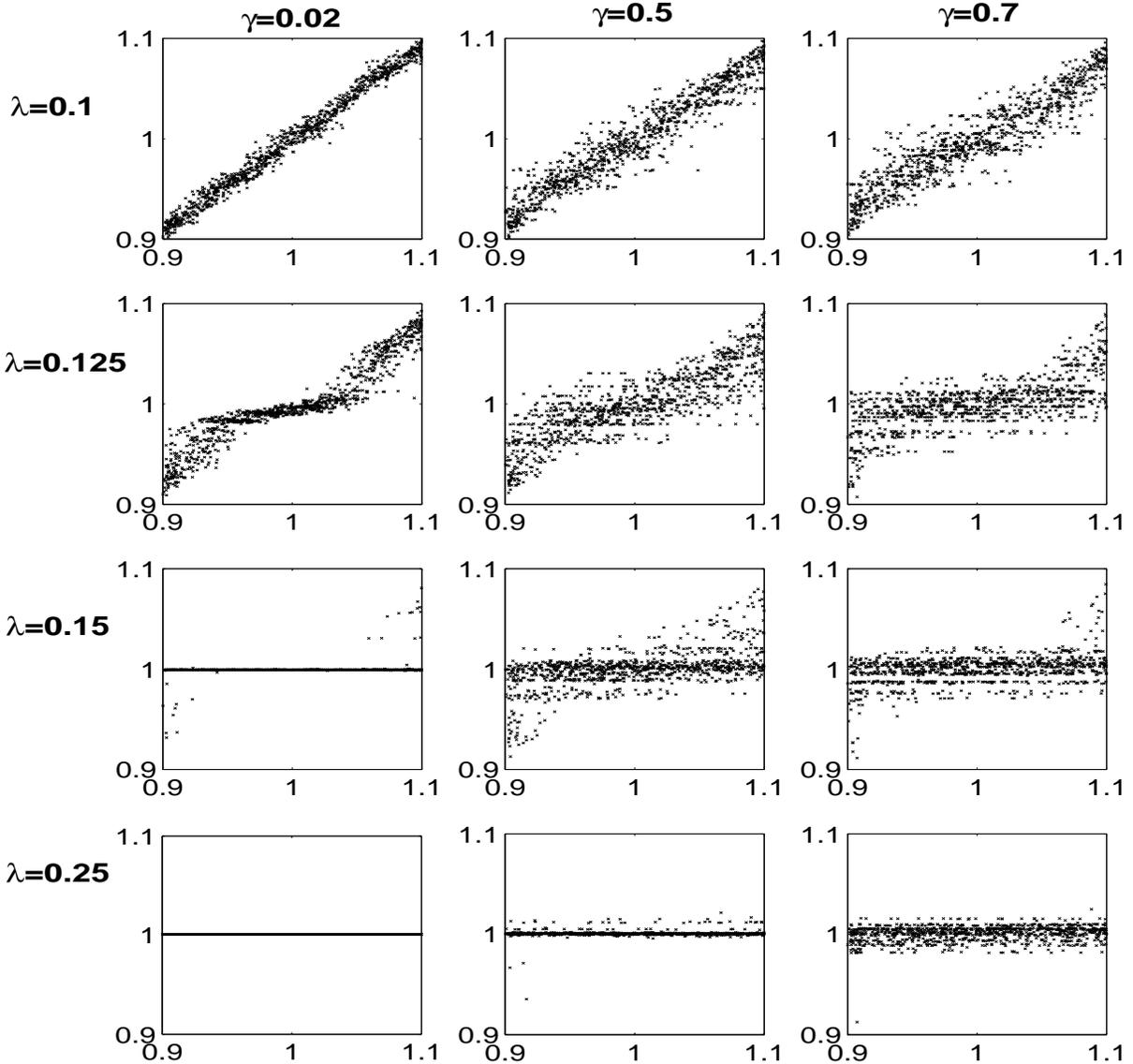}%
\caption{Poisson networks: Scatter plots of average oscillation frequency
$\Omega$ (vertical axis) vs. intrinsic frequency $\omega$ (horizontal). \ At
zero coupling, \ the points all lie on the line $\Omega=\omega$. \ With
increasing coupling, \ more oscillators line up at synchronized frequencies.
\ For high clustering, \ several bands form at different frequencies, rather
than one frequency as in the low-$\gamma$ case.}%
\label{poissonscatter}%
\end{center}
\end{figure}
%EndExpansion%
%TCIMACRO{\FRAME{ftbpFU}{6.3235in}{6.0226in}{0pt}{\Qcb{Scale-free networks:
%Scatter plots of $\Omega$ vs. $\omega$. \ Synchronized subsets (horizontal
%bands) begin to form at weaker coupling for the more clustered networks.
%\ Frequency clustering is evident at high $\gamma$.}}{\Qlb{sfscatter}%
%}{sfscatter.eps}{\special{ language "Scientific Word";  type "GRAPHIC";
%display "ICON";  valid_file "F";  width 6.3235in;  height 6.0226in;
%depth 0pt;  original-width 6.6668in;  original-height 5.2157in;
%cropleft "0";  croptop "1.0475";  cropright "0.9989";  cropbottom "0";
%filename 'sfscatter.eps';file-properties "XNPEU";}}}%
%BeginExpansion
\begin{figure}
[ptbptbptb]
\begin{center}
\includegraphics[
trim=0.000000in 0.000000in 0.007334in -0.247746in,
height=6.0226in,
width=6.3235in
]%
{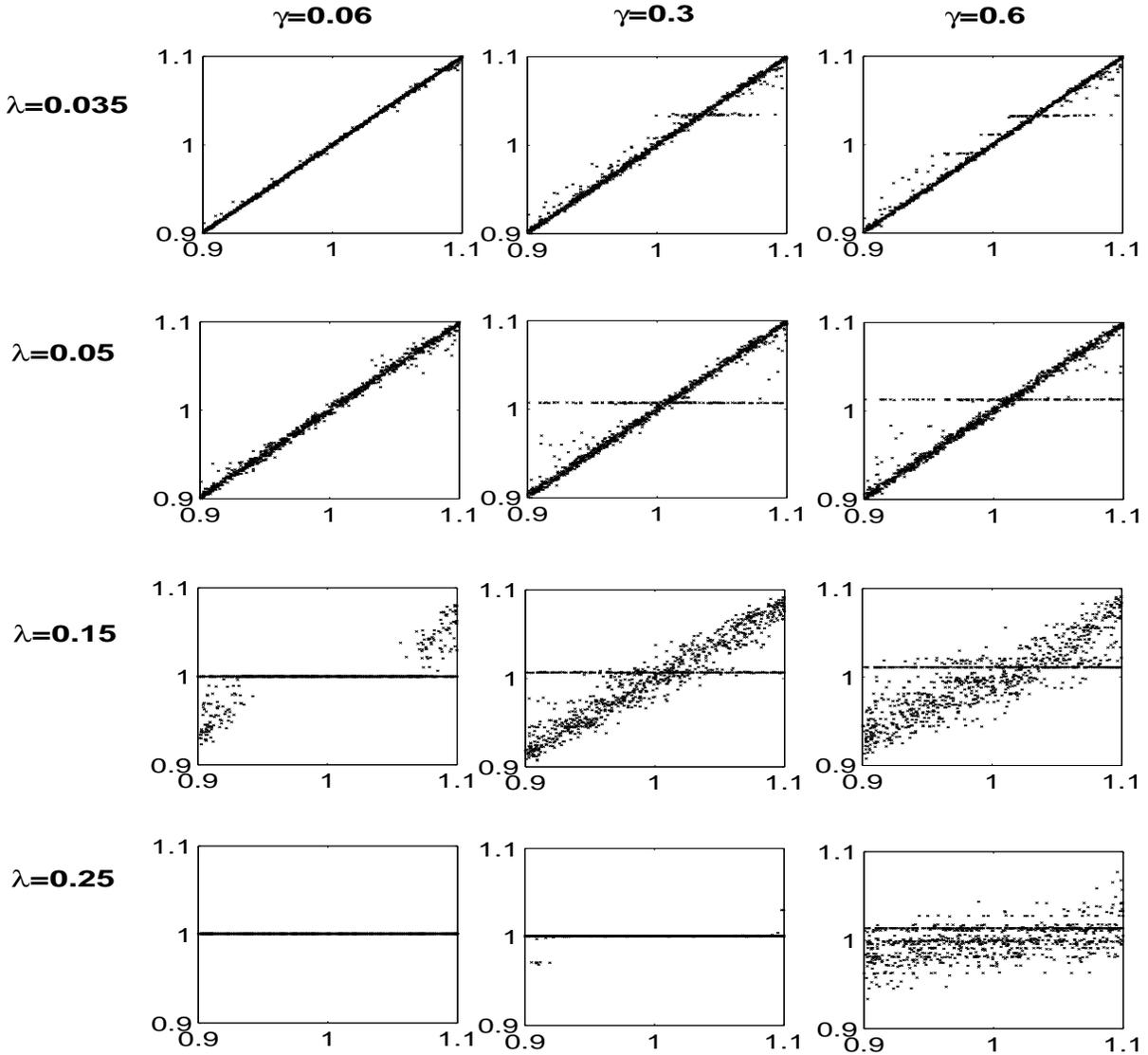}%
\caption{Scale-free networks: Scatter plots of $\Omega$ vs. $\omega$.
\ Synchronized subsets (horizontal bands) begin to form at weaker coupling for
the more clustered networks. \ Frequency clustering is evident at high
$\gamma$.}%
\label{sfscatter}%
\end{center}
\end{figure}
%EndExpansion%
%TCIMACRO{\FRAME{ftbpFU}{6.3235in}{6.0226in}{0pt}{\Qcb{Scale-free networks:
%Average frequency $\Omega$ vs degree $k$ at several values of clustering
%$\gamma$ and coupling strength $\lambda$. \ The degree (horizontal axis) is
%plotted on a logarithmic scale.\ Synchronization begins with the hubs (nodes
%with the highest $k$) and progresses downward to those with lower $k$.}%
%}{\Qlb{sfkscatter}}{sfkscatter.eps}{\special{ language "Scientific Word";
%type "GRAPHIC";  display "ICON";  valid_file "F";  width 6.3235in;
%height 6.0226in;  depth 0pt;  original-width 6.6789in;
%original-height 5.2157in;  cropleft "0";  croptop "1";  cropright "1";
%cropbottom "0";  filename 'SFKscatter.eps';file-properties "XNPEU";}}}%
%BeginExpansion
\begin{figure}
[ptbptbptbptb]
\begin{center}
\includegraphics[
height=6.0226in,
width=6.3235in
]%
{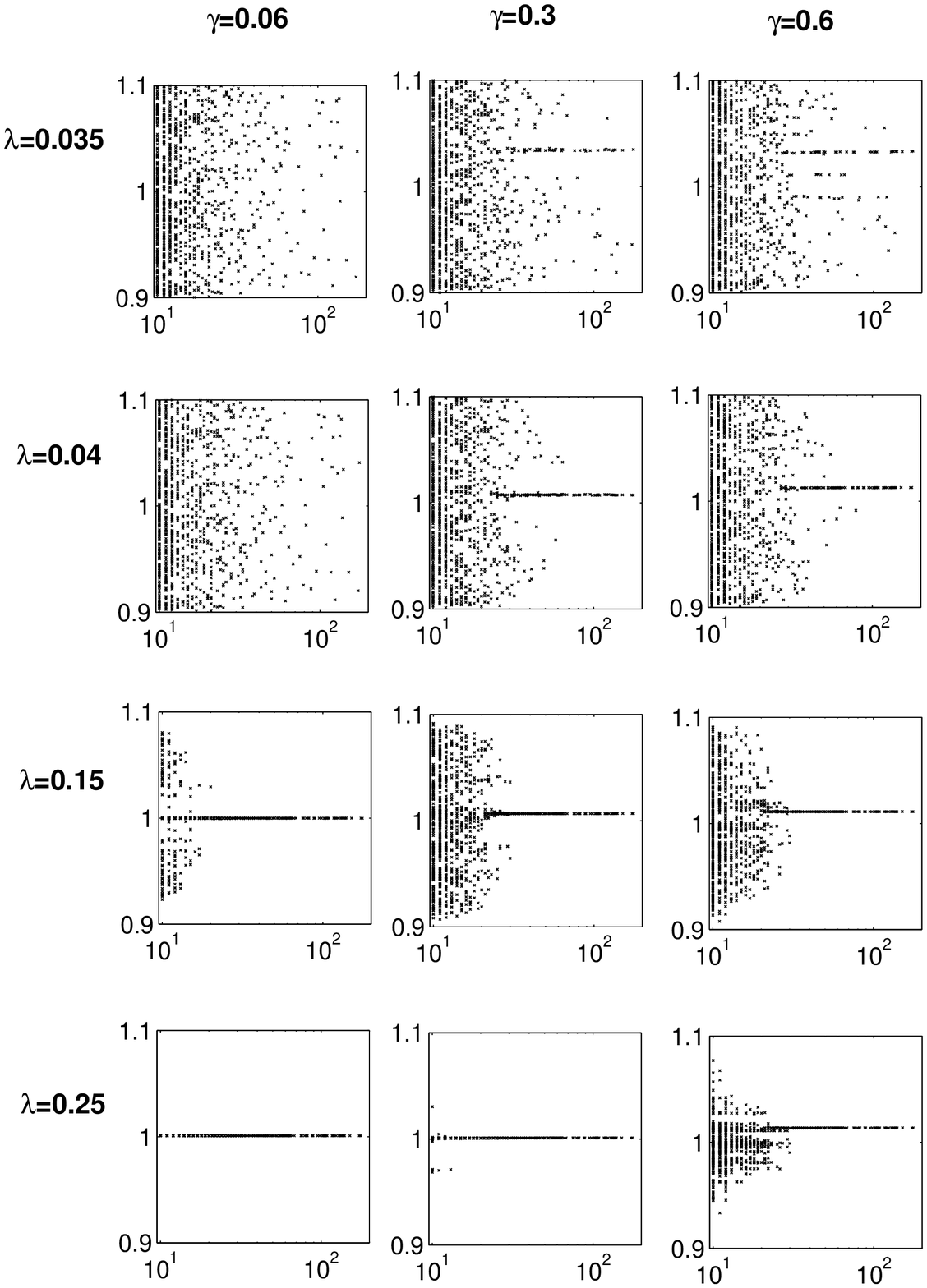}%
\caption{Scale-free networks: Average frequency $\Omega$ vs degree $k$ at
several values of clustering $\gamma$ and coupling strength $\lambda$. \ The
degree (horizontal axis) is plotted on a logarithmic scale.\ Synchronization
begins with the hubs (nodes with the highest $k$) and progresses downward to
those with lower $k$.}%
\label{sfkscatter}%
\end{center}
\end{figure}
%EndExpansion

In scale-free networks (fig.1B,C), clustering has a more complicated effect.
\ The network with $\gamma=0$ shows the onset of synchronization (upward turn
in the graph) near $\lambda\approx0.07$ and the transition to full
synchronization is not as steep as in the Poisson case. \ Surprisingly,
increased clustering slightly enhances the order parameter at weak coupling
but suppresses it at strong coupling. \ For intermediate $\gamma$ this results
in two separate transitions, an advanced transition (threshold at
$\lambda\approx0.03$) to partial synchronization followed by a delayed one
($\lambda\approx0.15$) to full synchronization. \ \ 

For more insight into the synchronization transitions we examine scatter plots
of observed frequency $\Omega_{i}$ vs. intrinsic frequency $\omega_{i}$.
\ Each plot in figures \ref{poissonscatter}-\ref{sfkscatter} shows the
population's behavior for one realization of the random variables. \ In the
absence of interactions, all points would fall along the line $\Omega
_{i}=\omega_{i}$. \ In the globally coupled (mean-field) case\cite{Kuramoto},
\ the nodes differ only by their intrinsic fequencies. \ \ Therefore, the
observed frequency is in every case a single-valued function $\Omega(\omega)$
of the intrinsic frequency, whose shape changes with $\lambda$. \ In
particular, the curves begin to flatten around $\omega\approx1$, as
oscillators with frequencies closest to the average are the first to
synchronize. \ \ The flat portion broadens with increasing coupling. \ \ For
Poisson networks (fig. \ref{poissonscatter}) at low clustering, the same
qualitative picture holds as in the mean-field case except that there is more
scatter and the curve is blurred. \ This is due to the fact that each
oscillator's behavior now depends not only on its intrinsic frequency and the
global average, but also on the details of its local neighborhood. \ As in the
mean-field case, the oscillators near the extremes of the $\omega$
distribution are the last to become fully entrained. \ Unlike the mean-field
case, \ some remain unentrained even when the synchronized group is
well-established and has recruited members from the entire intrinsic frequency
range. \ At higher clustering, the scatter plots differ even more from the
mean-field case, and they do so in two ways. \ First, \ the scatter is more
pronounced. \ Second, \ the horizontal striations indicate that the
oscillators cluster in subgroups (frequency clusters) oscillating at different
frequencies. \ As $\lambda$ increases, the frequency clusters converge and
eventually merge, \ but full synchronization requires a stronger coupling
compared to networks with low clustering. \ The beating of the different
frequencies accounts for the low value of the global order parameter and its
large fluctuations.\ \ 

In the scale-free case, \ the scatter plots (figure \ref{sfscatter}) confirm
that the highly clustered networks begin to synchronize at a weaker coupling
than the less clustered ones. \ Initially, however, this synchronization only
affects a subset of the nodes, while the remaining nodes still fall close to
the line $\Omega=\omega$ as if they were not interacting at all. \ This is in
contrast to the Poisson scatter plots which begin to flatten near the center
of the frequency range as the transition begins. \ Since scale-free networks
have a strongly heterogeneous degree distribution, the entrainment of an
oscillator depends strongly on the number of its inputs. \ When $\Omega$ is
plotted against the degree $k$ of each node as in figure \ref{sfkscatter},
\ it is apparent that the higher-degree nodes (hubs) begin to synchronize
first, \ while the lower-degree nodes synchronize as the coupling continues to
increase.\ \ One similarity with the Poisson case is that structural
clustering enhances frequency clustering. \ While clustering promotes the
formation of synchronized frequency clusters among the hubs, it inhibits the
synchronization of the network as a whole. \ The early hub synchronization
accounts for the slightly enhanced order parameter at weak coupling. \ 

In conclusion, we examined the synchronization of networks of non-identical
coupled phase oscillators with both Poisson and scale-free degree
distributions, \ and studied the effects of varying the clustering coefficient
without affecting the degree distribution. \ \ Our first main result, for both
types of networks, is that clustering encourages the formation of
sub-populations synchronized at different frequencies (frequency clusters),
\ and thus discourages full global synchronization at a single frequency.
\ \ Our second key finding concerns the scale-free case, where we found that,
despite the increased difficulty of \emph{\ full} synchronization, \ higher
clustering actually promotes the onset of \ \emph{partial} synchronization of
the hubs. \ Scale-free networks with high clustering thus appear to undergo
two separate transitions as the coupling strength increases: an early
transition to partial synchronization and a delayed one to full
synchronization. \ The first transition involves only the hubs while leaving
the majority almost unaffected. \ The hubs seem to form the growth nuclei for
the synchronized state. \ \ Interestingly, they are able to synchronize even
though many of their inputs come from lower degree nodes which are not yet
synchronized. \ It is known\cite{Moreno} that in ordinary, low-clustered
scale-free networks the relaxation time for synchronization of hubs is shorter
than that of less connected nodes, but this does not account for the effect of
structural clustering in promoting their synchronization. \ It is possible
that increasing clustering may, as a by-product, increase the assortativity of
degree mixing\cite{assortative}, so that hubs become more likely to connect to
each other. \ In other models of clustered networks it was found that
triangles preferentially include higher-degree nodes\cite{Sergi}. \ Clustering
may also affect the betweenness centrality distributions. \ 

This advanced transition may play a role in highly clustered networks, for
instance in the human brain. In natural as opposed to abstract networks, the
cost associated with long-range connections often gives them a tendency toward
clustering. \ For the ganglion of \textit{C. elegans} \cite{cherniak}, the
clustering coefficient is $\gamma=0.28$. For the human brain \cite{equiluz},
the connectivity is essentially scale-free in different regions, and
clustering lies three orders of magnitude above that of equivalent random
networks. This and the weak synaptic coupling strength in the brain
\cite{abeles}\cite{hoppenstaedt} point to the potential importance of
cluster-enhanced synchronization of oscillating hubs.

\end{document}